# An analytic model to calculate Voxel S-Values for $^{177}$Lu

**Daniele Pistone**[1,2], **Lucrezia Auditore**[1,2,*], **Antonio Italiano**[2,3], **Sergio Baldari**[1,4], **Ernesto Amato**[1,2]

[1] Department of Biomedical and Dental Sciences and of Morphologic and Functional Imaging (BIOMORF), University of Messina, Messina, Italy
[2] INFN, National Institute for Nuclear Physics, Section of Catania, Catania, Italy
[3] Department of Mathematical and Computational Sciences, Physics Sciences and Earth Sciences (MIFT), University of Messina, Messina, Italy
[4] Nuclear Medicine Unit, University Hospital "Gaetano Martino", Messina, Italy

**Corresponding author's email**: lauditore@unime.it

**Abstract**

*Objective*: $^{177}$Lu is one of the most employed isotopes in targeted radionuclide therapies and theranostics, and 3D internal dosimetry for such procedures has great importance. Voxel S-Values (VSVs) approach is widely used for this purpose, but VSVs are available for a limited number of voxel dimensions. The aim of this work is to develop an analytic model for the calculation of $^{177}$Lu-VSVs in any cubic voxelized geometry of practical interest.
*Approach*: Monte Carlo (MC) simulations were implemented with the toolkit GAMOS to evaluate VSVs in voxelized geometries of soft tissue from a source of $^{177}$Lu homogeneously distributed in the central voxel. Nine geometric setups, containing 15×15×15 cubic voxels of sides $l$ ranging from 2 mm to 6 mm, in steps of 0.5 mm, were considered. For each $l$, the VSVs computed as a function of the "normalized radius", $R_n = R/l$ (with $R$ = distance from the center of the source voxel), were fitted with a parametric function. The dependencies of the parameters as a function of $l$ were then fitted with appropriate functions, in order to implement the model for deducing $^{177}$Lu-VSVs for any $l$ within the aforementioned range.
*Main results*: The MC-derived VSVs were satisfactorily compared with literature data for validation, and the VSVs computed with the analytic model agree with the MC ones within 2% for $R_n \leq 2$ and within 6% for $R_n > 2$.
*Significance*: The proposed model enables the easy and fast calculation, with a simple spreadsheet, of $^{177}$Lu-VSVs in any cubic voxelized geometry of practical interest, avoiding the necessity of implementing *ad-hoc* MC simulations to estimate VSVs for specific voxel dimensions not available in literature data.

## 1. Introduction

Lutetium-177 is a β⁻ emitter with favorable nuclear characteristics as a therapeutic radionuclide, including its low energy β emissions ($E_{β\text{-max}}$ = 496.8 keV and $\langle E_β \rangle$ = 133.64 keV), its convenient physical half-life of 6.639 days and the low abundance of gamma emissions in its decays, useful for imaging (Kossert *et al* 2012, Pillai and Knapp 2015). Nuclear medicine therapies with Lutetium-labeled radiopharmaceuticals are expanding since, in addition to the traditional peptide receptor radionuclide therapy (PRRT) of neuroendocrine tumors (NETs) (Kim and Kim 2017, del Olmo-García *et al* 2022), $^{177}$Lu therapies with prostate specific membrane antigen (PSMA) ligands have been introduced for the treatment of prostate cancer (Emmett *et al* 2017, Sartor *et al* 2021), together with other Lutetium-labeled radiopharmaceuticals for the treatment of other diseases and for theranostic uses (Pillai and Knapp 2015, Das and Banerjee 2016).

The internal dosimetry of these therapies is necessary for the prediction of the therapeutic efficacy and for their safety; in this context, three-dimensional dosimetry is useful to estimate both the radiation absorbed doses in the target tissues of the therapy, and in the organs at risk (Berenato *et al* 2016, Marin *et al* 2018, Del Prete *et al* 2018, Ligonnet *et al* 2021, Sjögreen Gleisner *et al.* 2022).

The calculation approaches for three-dimensional internal dosimetry are: the convolution of dose point-kernels (DPKs), the convolution of voxel S-factors, also called voxel S-values (VSVs), and the direct Monte Carlo (MC) simulation (Amato *et al* 2022, Auditore *et al* 2022). Direct MC is the most accurate one, but requires demanding resources and is currently used only for research (Pistone *et al* 2021); therefore, DPKs and VSVs are the most widely used for clinical dosimetry (Dewaraja *et al* 2012). The VSVs convolution approach, introduced and described in detail in the MIRD pamphlet n.17 (Bolch *et al* 1999), has the advantage of not needing CPU–intensive conversion of spherical coordinates to Cartesian ones over the target volumes, differently from DPK convolution (Lee *et al* 2018). However, the VSVs approach requires tabulated S-values for the examined isotope with the corresponding voxel size of the considered tomographic scans, from which the time-integrated activity matrix for convolution is deduced. In fact, the different SPECT-CT scanners generally reconstruct with matrices of different sizes; for this reason, the available bibliographic resources, such as https://www.medphys.it/down_svoxel.htm, publish VSVs in a number of standard voxel sizes, which by the way is limited. For voxel dimensions not available in the literature, a viable but unpractical solution is to ask for a specific Monte Carlo calculation, to research groups active in the field. Alternatively, Amato *et al* (2012) proposed a general analytical method for the calculation of VSVs for beta- and gamma-emitting radionuclides. This method, however, is based on the interpolation of parameters in the two-dimensional space of energy of emitted radiation and voxel size. Fernández *et al* (2013) also tested analytic methods based on the down-sampling of high-resolution VSVs, on VSVs interpolation and/or fits, but did not provide tabular data nor analytic expressions in order to precisely reproduce their results, showing rather that such analytic methodologies are feasible and suggesting to the interested readers to develop them for specific radionuclides in future studies.

The purpose of this work is to introduce a specific analytic model for $^{177}$Lu, to be more accurate than the general method by Amato *et al* (2012) and which can be straightforwardly implemented in a simple electronic spreadsheet, in order to allow the computation of $^{177}$Lu VSVs in any cubic voxel dimension of practical interest.

## 2. Materials and Methods

### 2.1 Monte Carlo simulations

We developed Monte Carlo (MC) simulations for the calculation of VSVs for $^{177}$Lu using GAMOS (Arce *et al* 2008, Arce *et al* 2014), a GEANT4-based user-friendly framework for medical physics applications (Auditore *et al* 2019, Amato *et al* 2020, Pistone *et al* 2020). GEANT4 (Agostinelli *et al* 2003, Allison *et al* 2006, Allison *et al* 2016) is a simulation toolkit for radiation transport widely used and well validated in many fields of physics, including medical radiation physics and radioprotection, internal dosimetry, and VSVs calculations (Pacilio *et al* 2009, Amato and Lizio 2009, Amato *et al* 2013a, Amato *et al* 2013b).

In particular, we exploited GAMOS version 6.2.0, relying on GEANT4 version 10.06.p02. A cubic *World* volume of 50 cm side and made of soft tissue was set. In detail, G4_TISSUE_SOFT_ICRP material, with density 1.03 g·cm$^{-3}$ and elemental composition according to the ICRP definition (Geant4 Collaboration 2020) was adopted. Within the *World,* nine voxelized cubic geometries were considered, each one containing 15 × 15 × 15 voxels and centered in the origin of the reference system, with the following voxel sizes $l$: 2, 2.5, 3, 3.5, 4, 4.5, 5, 5.5 and 6 mm.

For each *l*, independent simulations were prepared, setting the central voxel as a homogeneous and isotropic source of $^{177}$Lu. The $^{177}$Lu decay was implemented from the data of (Stabin and da Luz 2002), publicly available at http://www.doseinfo-radar.com, performing a dedicated simulation for each of the three types of primary particles: β electrons, monoenergetic (Auger and Conversion) electrons, and X and γ photons, and then merging their results by accounting for their relative probabilities. The transport of the emitted radiation was simulated with the *GmEMExtendedPhysics* Physics List (Gamos Collaboration 2020), and the absorbed dose *per* event was scored in each of the aforementioned voxels, together with its respective statistical uncertainty in terms of standard deviation of the mean (Chetty *et al* 2006). These 3D dose outputs were then processed with home-made Python scripts in order to: *i*) label each voxel with indices (*i,j,k*) ranging from (-7,-7,-7) to (7,7,7); *ii*) convert the doses *per* event (Gy/event) into S-Values (mGy/Mbq·s); *iii*) average the S-Values of the symmetrical voxels with respect to the center of the volume, and calculate the statistical uncertainties; *iv*) produce text outputs reporting the S-Values in column as a function of the voxel indices and of the dimensionless normalized radius $R_n$, defined as:

$$R_n = \sqrt{i^2 + j^2 + k^2} = R/l \tag{1}$$

where *R* indicates the distance from the center of the central voxel.

For each run, $10^9$ histories were simulated, a number which guaranteed statistical uncertainties below 1% for the final S-Values, in all the voxels and for every *l* value. The simulations were run on a local workstation provided with *Intel(R) Core(TM) i7-10700K @ 3.80 GHz* CPUs and 32 GB RAM, and lasted on average 4 hours for β spectrum runs, 3 hours for monoenergetic electrons runs, and 9 hours for photons runs.

To validate the correct functioning of the simulations and post-processing procedures, the obtained S-Values, $S_{MC}(i,j,k)$, were compared with the ones of Lanconelli *et al* (2012) publicly available at https://www.medphys.it/down_svoxel.htm, $S_{Lan}(i,j,k)$, in terms of relative percent differences κ(*i,j,k*) (%) in each voxel, for the available corresponding *l* values (3, 4, 5 and 6 mm):

$$\kappa(i,j,k) = 100 \cdot \frac{S_{MC}(i,j,k) - S_{Lan}(i,j,k)}{S_{Lan}(i,j,k)} \tag{2}$$

The $S_{Lan}(i,j,k)$ had been calculated via direct MC simulation, using the EGSnrc-based DOSXYZnrc code, and were in turn successfully compared with the results of two other MC codes: MCNP4c and PENELOPE (Lanconelli *et al* 2012).

**2.2 Analytic model**

The following function was selected to fit the MC-derived S-Values represented as a function of $R_n$, $S_{MC}(R_n)$:

$$S(R_n) = a \cdot exp(-exp(b \cdot R_n^c)) + \frac{f}{(R_n^g + 0.02)} \tag{3}$$

where *a, b, c, f* and *g* are the fit parameters for each examined *l*. The first term in the right side of Eq. 3 is derived by the empirical expression introduced in Eq. 4 of Amato *et al* (2012) to model $S(R_n)$ for monoenergetic electrons, and in our work it is aimed at describing the contribution of β-particles and monoenergetic electrons of $^{177}$Lu decay to the S-Values. The second term of Eq. 3 is derived by Eq. 5 of Amato *et al* (2012) and it is aimed at describing the contribution of the photons of $^{177}$Lu decay to the S-Values.

The five parameters of Eq. 3 were then fitted as a function of *l*; *a* and *f* were fitted with a function of the following type:

$$y(l) = \frac{p_0}{(l^{p_1} + p_2)} \tag{4}$$

with $p_0$, $p_1$ and $p_2$ as fit parameters. *b, c* and *g* were fitted with a 3$^{rd}$-order polynomial function:

$$y(l) = q_0 + q_1 l + q_2 l^2 + q_3 l^3 \tag{5}$$

with $q_0$, $q_1$, $q_2$ and $q_3$ as fit parameters.

As it will be clear from the results, a corrective term for the unique $S(1,1,1)$ located at $R_n \simeq 1.732$ had to be introduced, $\delta(l)$:

$$S(1,1,1,l) = S_{Eq.3}(1,1,1,l) + \delta(l) \qquad (6)$$

$\delta(l)$ was obtained by fitting the differences between the MC-derived S-Value, $S_{MC}(1,1,1,l)$, and the S-Value deduced from the fit of Eq. 3, $S(1,1,1,l)$; a bi-exponential fit function was adopted, since it demonstrated to optimally reproduce the trend of the differences as a function of $l$:

$$\delta(l) = A1 \cdot exp(-l/t1) + A2 \cdot exp(-l/t2) + y0 \qquad (7)$$

with $A1$, $A2$, $t1$, $t2$ and $y0$ as parameters.

All the fits described in this Section were performed with the software QtiPlot[1] 0.9.8.9 using the Nelder-Mead Simplex algorithm.

The analytic model constituted by Eqs. 3, 4 and 5 and including the corrective term for voxel (1,1,1) of Eqs. 6 and 7 was implemented in a simple spreadsheet, enabling the calculation of VSVs for $^{177}$Lu for whatever voxel dimension $l$ between 2 mm and 6 mm.

In order to validate the analytic model, the VSVs calculated through it were first compared with the MC-derived ones for all the $l$ values for which they were simulated (reported in Sec. 2.1). Secondly, they were compared for three additional $l$ values randomly selected between 2 mm and 3 mm, 3 mm and 5 mm, and 5 mm and 6 mm, to assess the capability of the model of calculating correctly VSVs for small, medium and large voxel sizes of practical interest. For these three $l$ values, namely 2.68 mm, 4.32 mm and 5.35 mm, MC simulations were performed to evaluate S-Values in the same way described in Sec. 2.1, and were compared with the S-Values from the analytic model. All the mentioned comparisons were reported in terms of relative percent differences $\varepsilon(i,j,k)$ (%) defined as follows:

$$\varepsilon(i,j,k) = 100 \cdot \frac{S(i,j,k) - S_{MC}(i,j,k)}{S_{MC}(i,j,k)} \qquad (8)$$

In order to assess how the accuracy of our new analytic model compares with the existing general model by Amato *et al* (2012), the VSVs calculated according to Amato *et al* (2012) were also compared in terms of $\varepsilon$ with the MC-derived ones.

As an additional feature of the model, the possibility to apply a density correction to the analytic S-Values following the approach of Dieudonné *et al* (2013) and Kim *et al* (2022) was included in the spreadsheet implementing the calculation, so that the interested users can adjust the density in the VSVs calculation. The density corrected S-Values, $S_\rho(R_n)$, are calculated as:

$$S_\rho(R_n) = S(R_n) \frac{1.03}{\rho} \qquad (9)$$

where $\rho$ is the user-defined density (in g·cm$^{-3}$) and 1.03 g·cm$^{-3}$ is the density of the soft tissue for which the MC VSVs were calculated (Sec. 2.1).

## 3. Results

**3.1 Monte Carlo simulations validation**

In Fig. 1 are reported the relative percent differences $\kappa$ (Eq. 2) between the S-Values obtained with the MC simulations performed in this work and the ones by Lanconelli *et al* (2012), as a function of $R_n$; the highest value of normalized radius is $R_n \simeq 8.66$, corresponding to the voxel (5,5,5), since it is the farthest one from the origin in the data selected for the comparison, which used voxel grids of 11 × 11 ×11. All the $\kappa$ values lie within ±2% except for $R_n$ = 0, i.e. for $S(0,0,0)$, for which $\kappa \simeq +9\%$.

---
[1] https://www.qtiplot.com

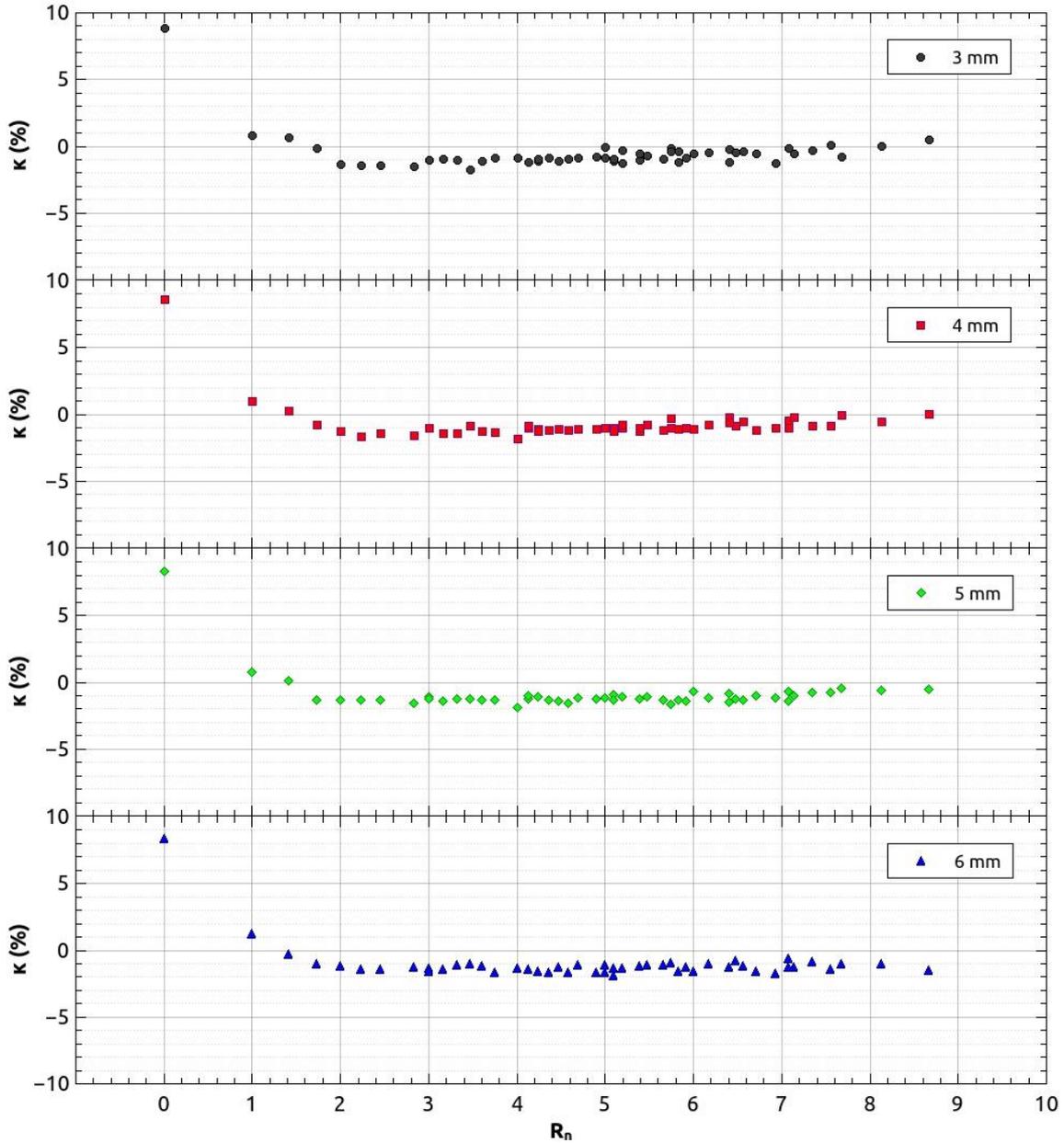

**Fig. 1** Relative percent differences $\kappa$ (Eq. 2) as a function of $R_n$ for the $l$ values available for comparison: 3, 4, 5 and 6 mm.

**3.2 VSVs from Monte Carlo simulations and analytic model**
In Figures 2 and 3, the Voxel S-Values, evaluated with the GAMOS MC simulations, are plotted as a function of $R_n$ (listed in tabular form in the Supplementary data), together with the fits performed using the analytic model function of Eq. 3; in the bottom panels the relative percent differences $\varepsilon(i,j,k)$ (Eq. 8) are reported, evaluated for each discrete $R_n$ value of the MC VSVs. The VSVs, fits and $\varepsilon$'s for the different voxel dimensions were splitted into two figures (Fig. 2 for 2, 3, 4, 5 and 6 mm, Fig. 3 for 2.5, 3.5, 4.5 and 5.5 mm) for clarity purpose, in order to avoid graphical superpositions of markers and curves. Error bars were omitted since the uncertainties were below 1% for all the data points, even at the farthest distances from the source.

All the fits of the VSVs converged with $R^2 > 0.99$, and the values of the obtained parameters are listed in Tab. A1 of the Annex, and are also represented as a function of $l$ in Fig. 4. Fig. 4 also shows the fits of the mentioned parameters as a function of $l$ with the functions reported in Eqs. 4 and 5, which all converged with $R^2 > 0.99$; the optimized parameters of these fits are reported in Tabs. A2 and A3 of the Annex and were used to build the spreadsheet implementing the analytic model, available in the Supplementary data and including the possibility to apply density correction, according to Eq. 9.

**3.3 Validation of the analytic model**

Considering the comparison between $S_{MC}$ and $S$ from the analytic model, despite the excellent goodness of the fits, significant discrepancies were found for the single $S(1,1,1)$, with $\varepsilon(1,1,1)$ values up to about -31%, depending on $l$, whereas all the other $\varepsilon(i,j,k)$ remained within ±6%, and within ±2% for $R_n < 2$.

Adopting the corrective term introduced in Eqs. 6 and 7, the updated $\varepsilon(1,1,1)$ values lie well below ±1%, as noticeable in the lower panel of Figs. 2 and 3, where they are depicted as open markers. The corrective factor $\delta(l)$ as a function of $l$ is shown in Fig. 5, and the fit parameters of $\delta(l)$ are reported in Tab. A4 of the Annex.

Considering the comparison of MC simulations and analytic model for the three test voxel sizes 2.68 mm, 4.32 mm and 5.35 mm (Sec. 2.2), in Fig. 6 are reported their $S_{MC}(R_n)$ and $S(R_n)$ including the correction term for $S(1,1,1)$, together with their respective $\varepsilon(i,j,k)$, lying within -6% and +3% for $l = $ 5.35 mm and within ±2% for $l = $ 2.68 mm and 4.32 mm. The MC-derived $S_{MC}(R_n)$ for these three voxel dimensions are also reported in tabular form in the Supplementary data.

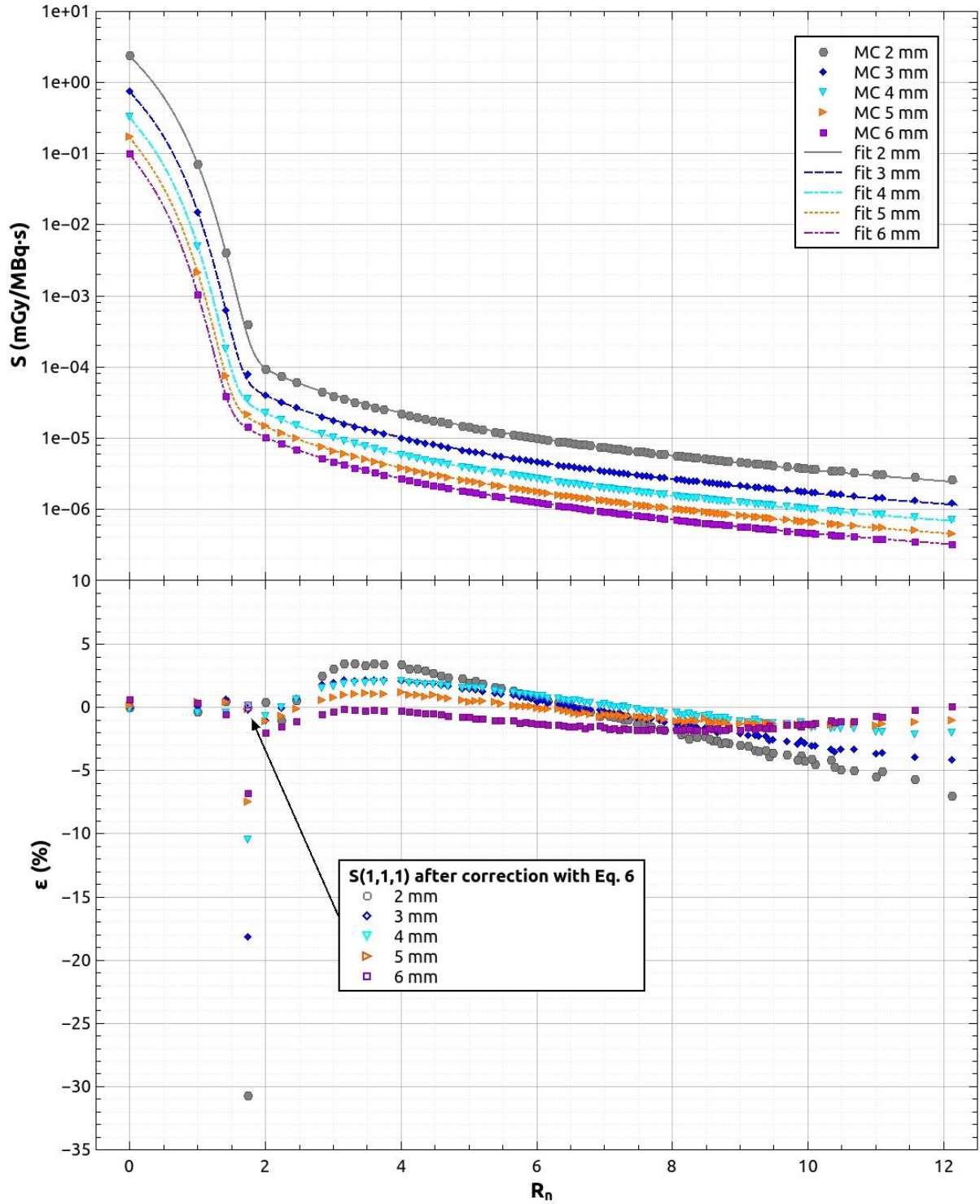

**Fig. 2** Top panel: S-Values calculated with GAMOS MC simulations for $l$ = 2, 3, 4, 5 and 6 mm, and fits with the function of Eq. 3; bottom panel: corresponding ε values (Eq. 8), with open markers for $R_n \simeq 1.732$ indicating the updated values after the correction of Eq. 6, which reduces ε(1,1,1) to less than 1%.

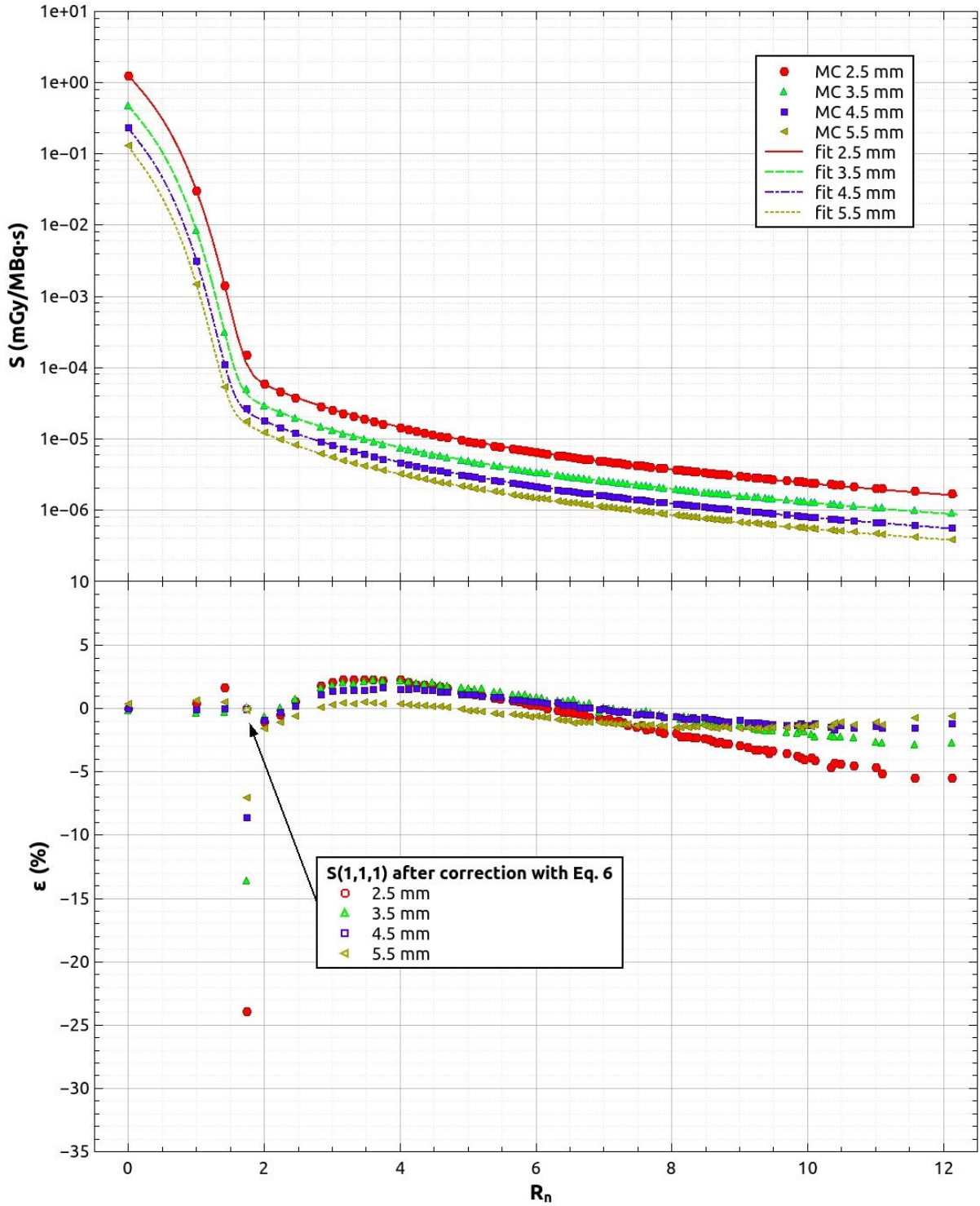

**Fig. 3** Top panel: S-Values calculated with GAMOS MC simulations for $l$ = 2.5, 3.5, 4.5 and 5.5 mm, and fits with the function of Eq. 3; bottom panel: corresponding ε values (Eq. 8), with open markers for $R_n \simeq 1.732$ indicating the updated values after the correction of Eq. 6, which reduces ε(1,1,1) to less than 1%.

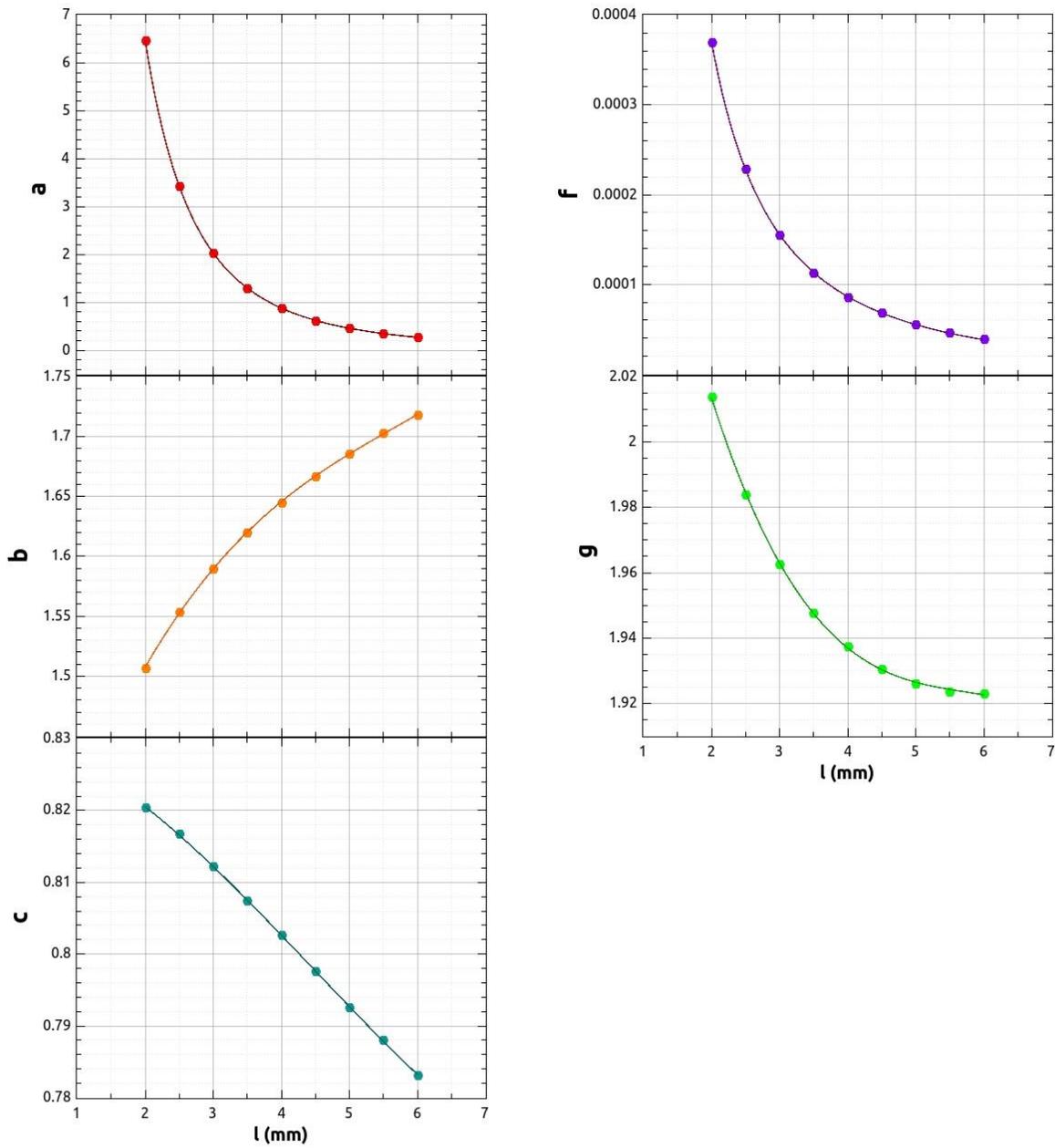

**Fig. 4** Values (markers) of the parameters *a, b, c, f* and *g* of Eq. 3. as a function of *l* and respective fits (lines) with Eq. 4 for *a* and *f*, with Eq. 5 for *b*, *c* and *g*.

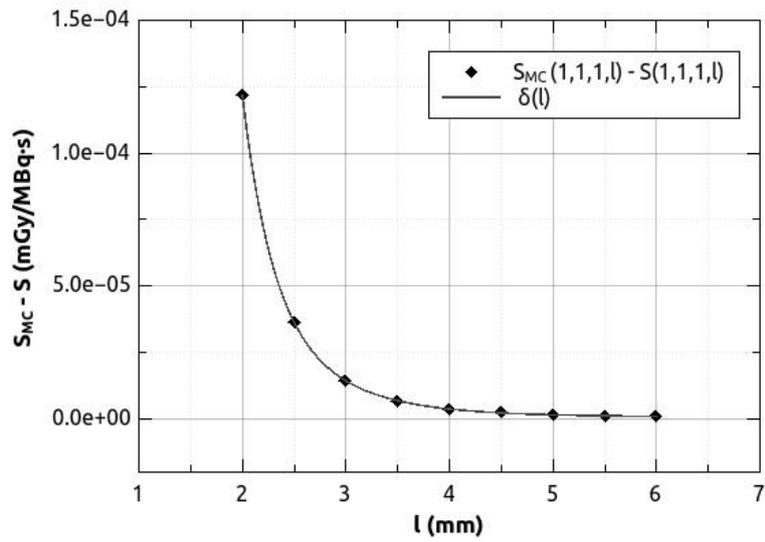

**Fig. 5** Differences between S-Values calculated with GAMOS MC simulations and with the fit function of Eq. 3 for the voxel (1,1,1), and fit $\delta(l)$ as a function of $l$ according to Eq. 7.

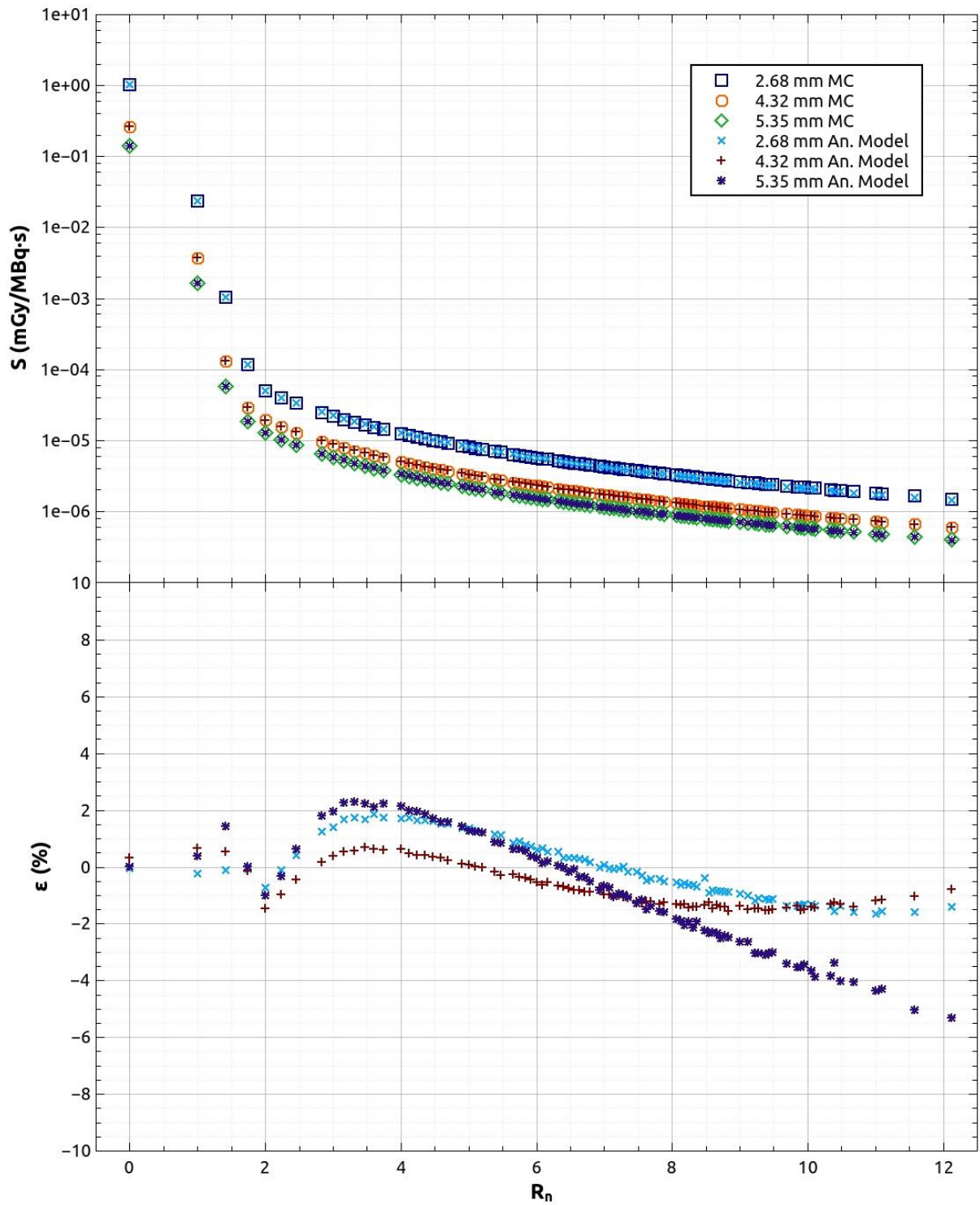

**Fig. 6** Top panel: S-Values calculated with GAMOS MC simulations and with the analytic model for voxel dimensions $l$ = 2.68 mm, 4.32 mm and 5.35 mm; bottom panel: corresponding ε values (Eq. 8).

# 4. Discussion

As Figs. 2 and 3 show, all the MC-derived $^{177}$Lu-VSVs exhibit a similar behavior as a function of $R_n$ for the different $l$'s: *i)* an early rapid decrease for $R_n < 2$, corresponding to the energy deposition from β's and monoenergetic electrons; *ii)* a smooth knee of the trend at $R_n \sim 2$, in correspondence of the maximum range of $^{177}$Lu β's in soft tissue (1.7 mm, according to Hosono *et al* 2018); *iii)* a slower decrease for $R_n > 2$, corresponding to the contribution of the monoenergetic photons of $^{177}$Lu decay and also to Bremsstrahlung photons produced by interactions of electrons in the medium. The similar trend of all the VSVs justifies the use of the function of Eq. 3 to fit them. This feature in addition translates in a very smooth variation with $l$ of all the fit parameters, exhibiting monotonically increasing or monotonically decreasing trends, and supporting the reliability of the fitting method over the range of voxel dimensions considered.

Concerning the validation of the MC simulations done through the comparison with the literature data of Lanconelli *et al* (2012), it appears largely satisfactory for $R_n > 0$, being the relative percent differences $\kappa$ within 2% for all the $l$ values available. The larger differences for $R_n = 0$, with $\kappa \simeq 9\%$, aside from reasonable discrepancies caused by the use of different simulation software and settings, is most likely due to the fact that in Lanconelli *et al* (2012) no mention is made of monoenergetic electrons of $^{177}$Lu decay, which probably were neglected in that work. This hypothesis is supported by the fact that we found out that the relative contribution of monoenergetic electrons to the total absorbed dose in the central voxel is of about 10%, which perfectly matches the missing 9% of Lanconelli's VSVs with respect to ours.

The analytic model developed in this work for the calculation of $^{177}$Lu-VSVs was satisfactorily validated comparing its results with MC ones for all the examined voxel dimensions: the relative percent differences ε are within ±6%, and by the way these discrepancies are found for $R_n$ values at the end of the tails of the VSVs, corresponding to the farthest voxels from the source, with the smallest contribution to the total absorbed dose; for $R_n < 2$, i.e. within the range of $^{177}$Lu β's, for which a prominent contribution to the absorbed dose is given, ε's are always within ±2%, including $S(1,1,1)$ after the incorporation of the voxel-specific correction term. These results demonstrated a significantly improved accuracy with respect to the pre-existing analytic model by Amato *et al* (2012), whose ε's s are within –20% and 0% for $R_n < 2$ and between –20% and +40% overall, as shown in Fig. 7.

In the work of Fernández *et al* (2013) it was suggested that analytic models based on the fitting of VSVs for specific radionuclides are feasible and their development is encouraged. In particular, they reported good results using polynomials fitting functions for separate intervals of voxel sizes; however, the order of the polynomials and the fit parameters were not reported in their work. In the present work, a unique function for fitting the entire range of examined voxel sizes is provided (Eq. 3), together with the functions for fitting the involved parameters as a function of the voxel size (Eq. 4 and 5) plus a corrective term for $S(1,1,1)$ (Eq. 6 and 7). In addition, our results are provided in the Annex in tabular form, and a spreadsheet automatically implementing all the mentioned calculations is provided as Supplementary material.

In view of all the described features, the proposed analytic method shows to be a solid radionuclide-specific calculation tool for $^{177}$Lu-VSVs. It enables an accurate, very simple and fast calculation of VSVs, requiring from the user only the input of the desired voxel dimensions in the provided spreadsheet, which instantaneously calculates all the parameters and consequently the VSVs, according to the introduced model. At will, also a desired density for the VSVs estimation can be given as input, to obtain density-corrected VSVs. Apart from the user-friendliness and computation rapidity, a major strength of the model is its capability of enabling the calculation of $^{177}$Lu-VSVs for whatever decimal voxel dimension between 2 mm and 6 mm. When dealing with 3D activity images deduced from PET or SPECT matrices with voxel dimensions not exactly matching the standard values commonly found in literature, this model permits to perform VSVs-based dosimetry avoiding the implementation of *ad-hoc* MC simulations to calculate the VSVs for specific voxel dimensions, while maintaining an high accuracy in the dosimetric calculation.

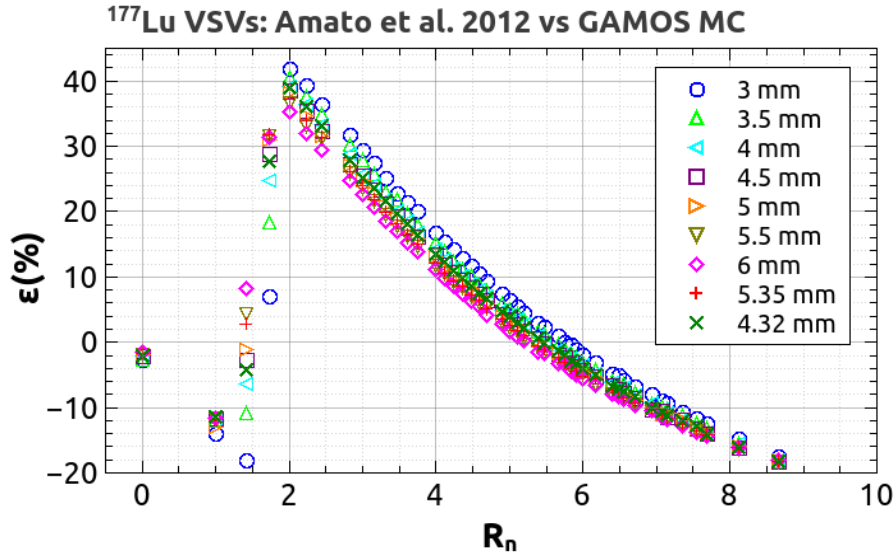

**Fig. 7** Relative percent differences ε between the VSVs by Amato *et al* (2012) and the MC-derived ones of the present work, for all the available voxel dimensions (the model by Amato *et al* (2012) was built for voxel sizes larger than 3 mm).

## 5. Conclusion

The proposed analytic model allows the easy and fast calculation of VSVs for $^{177}$Lu in any cubic voxel dimension between 2 mm and 6 mm by means of a simple spreadsheet. It ensures VSVs estimations in agreement with MC ones with ε < 2% for normalized radii within the maximum range of $^{177}$Lu β-electrons, and ε < 6% in the farthest voxels from the central one. This approach permits to perform VSVs-based dosimetry employing any tomographic activity matrix of practical interest, avoiding the need of implementing *ad-hoc* MC simulations to deduce VSVs for specific voxel dimensions.

**References**

Agostinelli S *et al* 2003 Geant4 - A simulation toolkit. *Nucl Instrum Methods Phys Res A.* **506** 250-303. https://doi.org/10.1016/S0168-9002(03)01368-8.
Allison J *et al* 2006 Geant4 Developments and Applications. *IEEE Trans Nucl Sci.* **53** 270-8. https://doi.org/10.1109/TNS.2006.869826.
Allison J *et al* 2016 Recent Developments in Geant4. *Nucl Instrum Methods Phys Res A.* **835** 186-225. https://doi.org/10.1016/j.nima.2016.06.125.
Amato E. and Lizio D 2009 Plastic materials as a radiation shield for β- Sources: A comparative study through Monte Carlo calculation. *J. Radiol. Prot.* **29** 239-50. https://doi.org/10.1088/0952-4746/29/2/010.
Amato E, Minutoli F, Pacilio M, Campennì A and Baldari S 2012 An analytical method for computing voxel S values for electrons and photons. *Med. Phys.* **39** 6808-6817. https://doi.org/10.1118/1.4757912
Amato E, Italiano A, Minutoli,F., Baldari S 2013a Use of the GEANT4 Monte Carlo to determine three-dimensional dose factors for radionuclide dosimetry, *Nucl Instrum Methods Phys Res A: Accelerators, Spectrometers, Detectors and Associated Equipment*, **708** 15-18 https://doi.org/10.1016/j.nima.2013.01.014.
Amato E, Italiano A and Baldari S 2013b Monte Carlo study of voxel S factor dependence on tissue density and atomic composition. *Nuclear Instruments and Methods in Physics Research A* **729** 870-876. https://doi.org/10.1016/j.nima.2013.08.059.
Amato E *et al* 2020 Full Monte Carlo internal dosimetry in nuclear medicine by means of GAMOS. *Journal of Physics: Conference Series.* **1561** 012002. https://doi.org/10.1088/1742-6596/1561/1/012002.
Amato E, Gnesin S, Cicone F and Auditore L 2022 Fundamentals of internal radiation dosimetry. *Reference Module in Biomedical Sciences*, Elsevier, ISBN 9780128012383, https://doi.org/10.1016/B978-0-12-822960-6.00142-3.

Arce P, Rato P, Cañadas M and Lagares JI 2008 GAMOS: A GEANT4-based easy and flexible framework for nuclear medicine applications. *2008 IEEE Nucl Sci Symp Conf Rec*, 3162-68. https://doi.org/10.1109/NSSMIC.2008.4775023.

Arce P *et al* 2014 Gamos: A framework to do Geant4 simulations in different physics fields with an user-friendly interface. *Nucl Instrum Methods Phys Res A* **735** 304–13. https://doi.org/10.1016/j.nima.2013.09.036.

Auditore L *et al* 2019 Internal dosimetry for TARE therapies by means of GAMOS Monte Carlo simulations. *Phys Med* **64** 245–51. https://doi.org/10.1016/j.ejmp.2019.07.024.

Auditore L, Pistone D, Amato E and Italiano A 2022 Monte Carlo methods in nuclear medicine. *Reference Module in Biomedical Sciences*, Elsevier, ISBN 9780128012383, https://doi.org/10.1016/B978-0-12-822960-6.00136-8.

Berenato S, Amato E, Fischer A and Baldari S 2016 Influence of voxel S factors on three-dimensional internal dosimetry calculations. *Physica Medica* **32**(10) 1259–1262. https://doi.org/10.1016/j.ejmp.2016.09.012

Bolch WE *et al* 1999 MIRD pamphlet no. 17: the dosimetry of nonuniform activity distributions—radionuclide S values at the voxel level. *J Nucl Med*. **40**(suppl) 11S–36S.

Chetty IJ *et al* 2006 Reporting and analyzing statistical uncertainties in Monte Carlo-based treatment planning *Int. J. Radiat. Oncol. Biol. Phys.* **65** 1249–59, https://doi.org/10.1016/j.ijrobp.2006.03.039

Das T and Banerjee S 2016 Theranostic Applications of Lutetium-177 in Radionuclide Therapy. *Curr Radiopharm.* **9**(1) 94-101. https://doi.org/10.2174/1874471008666150313114644

del Olmo-García MI, Prado-Wohlwend S, Bello P, Segura A and Merino-Torres JF 2022 Peptide Receptor Radionuclide Therapy with [177Lu]Lu-DOTA-TATE in Patients with Advanced GEP NENS: Present and Future Directions. *Cancers* **14**(3) 584. https://doi.org/10.3390/cancers14030584

Del Prete M *et al* 2018 Accuracy and reproducibility of simplified QSPECT dosimetry for personalized 177Lu-octreotate PRRT. *EJNMMI Phys* **5** 25 https://doi.org/10.1186/s40658-018-0224-9

Dewaraja YK *et al* 2012 MIRD pamphlet no. 23: Quantitative SPECT for patient-specific 3-dimensional dosimetry in internal radionuclide therapy *Journal of Nuclear Medicine* **53**(8) 1310–1325. https://doi.org/10.2967/jnumed.111.100123.

Dieudonné A *et al* 2013 Study of the impact of tissue density heterogeneities on 3-dimensional abdominal dosimetry: comparison between dose kernel convolution and direct Monte Carlo methods *J Nucl Med* **54** 236–243 https://doi.org/10.2967/jnumed.112.105825

Emmett L *et al* 2017 Lutetium 177 PSMA radionuclide therapy for men with prostate cancer: a review of the current literature and discussion of practical aspects of therapy. *Journal of medical radiation sciences* **64**(1) 52–60. https://doi.org/10.1002/jmrs.227

Fernández M *et al* 2013 A fast method for rescaling voxel S values for arbitrary voxel sizes in targeted radionuclide therapy from a single Monte Carlo calculation *Med Phys* **40** 082502 https://doi.org/10.1118/1.4812684

Gamos Collaboration 2020, GAMOS User's Guide Release 6.2.0, http://fismed.ciemat.es/GAMOS/GAMOS_doc/GAMOS.6.2.0/html/GamosUsersGuide_V6.2.0.html [last accessed May 2022].

Geant4 Collaboration 2020, Book For Application Developers v10.06.p02 Rev4.1, https://geant4-userdoc.web.cern.ch/UsersGuides/ForApplicationDeveloper/BackupVersions/V10.6c/html/index.html [last accessed May 2022].

Hosono M *et al* 2018 Manual on the proper use of lutetium-177-labeled somatostatin analogue (Lu-177-DOTA-TATE) injectable in radionuclide therapy (2nd ed.). *Ann Nucl Med.* **32**(3) 217-235. https://doi.org/10.1007/s12149-018-1230-7

Kim K and Kim SJ 2018 Lu-177-Based Peptide Receptor Radionuclide Therapy for Advanced Neuroendocrine Tumors. *Nucl Med Mol Imaging.* **52**(3) 208-215. https://doi.org/10.1007/s13139-017-0505-6

Kim KM *et al* 2022 Comparison of voxel S-value methods for personalized voxel-based dosimetry of 177 Lu-DOTATATE *Med Phys* **49**(3) 1888-1901 https://doi.org/10.1002/mp.15444

Kossert K, Nähle OJ, Ott O and Dersch R 2012 Activity determination and nuclear decay data of 177Lu. *Appl Radiat Isot.* **70**(9) 2215-2221. https://doi.org/10.1016/j.apradiso.2012.02.104.

Lanconelli N *et al* 2012. A free database of radionuclide voxel S values for the dosimetry of nonuniform activity distributions. *Physics in medicine and biology.* **57** 517-33. https://doi.org/10.1088/0031-9155/57/2/517

Lee MS et al. 2018 Whole-Body Voxel-Based Personalized Dosimetry: The Multiple Voxel S-Value Approach for Heterogeneous Media with Nonuniform Activity Distributions. *J Nucl Med.* **59**(7) 1133-1139. https://doi.org/10.2967/jnumed.117.201095

Ligonnet T *et al* 2021 Simplified patient-specific renal dosimetry in 177Lu therapy: a proof of concept. *Phys Med.* **92** 75-85. https://doi.org/10.1016/j.ejmp.2021.11.007

Marin G *et al* 2018 A dosimetry procedure for organs-at-risk in 177Lu peptide receptor radionuclide therapy of patients with neuroendocrine tumours. *Phys Med* **56** 41–9. https://doi.org/10.1016/j.ejmp.2018.11.001.

Pacilio M *et al* 2009 Differences among Monte Carlo codes in the calculations of voxel S-values for radionuclide targeted therapy and analysis of their impact on absorbed dose evaluations. *Med. Phys.* **36** 1543-1552. https://doi.org/10.1118/1.3103401

Pillai AM and Knapp FF 2015 Evolving Important Role of Lutetium-177 for Therapeutic Nuclear Medicine. *Curr Radiopharm.* **8**(2) 78-85. https://doi.org/10.2174/1874471008666150312155959

Pistone D *et al* 2020 Monte Carlo based dose-rate assessment in 18F-choline pet examination: A comparison between gate and gamos codes. *AAPP Atti Della Accademia Peloritana dei Pericolanti - Classe di Scienze Fisiche, Matematiche e Naturali,* **98**(A5) 1–15. https://doi.org/10.1478/AAPP.981A5

Pistone D *et al* 2021 GATE Monte Carlo dosimetry in 90Y TARE planning: influence of simulation parameters and image resampling on dosimetric accuracy and optimization of computational times. *AAPP Atti della Accademia Peloritana dei Pericolanti - Classe di Scienze Fisiche, Matematiche e Naturali,* **99**(A4) 1–35. https://doi.org/10.1478/AAPP.992A4

Sartor O *et al* 2021 Lutetium-177-PSMA-617 for Metastatic Castration-Resistant Prostate Cancer. *The New England journal of medicine* **385**(12) 1091–1103. https://doi.org/10.1056/NEJMoa2107322

Sjögreen Gleisner K *et al* 2022 EANM dosimetry committee recommendations for dosimetry of 177Lu-labelled somatostatin-receptor- and PSMA-targeting ligands. *Eur J Nucl Med Mol Imaging* **49** 1778–1809 https://doi.org/10.1007/s00259-022-05727-7

Stabin, MG and da Luz LCQP 2002 Decay data for internal and external dose assessment, *Health Phys* **83**(4) 471-475, https://doi.org/10.1097/00004032-200210000-00004

**Annex**

**Tab. A1** Values of the fit parameters, $\chi^2$/dof and $R^2$ for the fits of the $S_{MC}(R_n)$ with Eq. 3 of this work.

| $l$ (mm) | $a$ | $b$ | $c$ | $f$ | $g$ | $\chi^2$/dof | $R^2$ |
|---|---|---|---|---|---|---|---|
| 2.0 | 6.4692E+00 | 1.5069E+00 | 8.2044E-01 | 3.7025E-04 | 2.0139E+00 | 4.0710E-05 | 0.99998 |
| 2.5 | 3.4290E+00 | 1.5533E+00 | 8.1672E-01 | 2.2860E-04 | 1.9841E+00 | 1.0296E-05 | 0.99999 |
| 3.0 | 2.0302E+00 | 1.5899E+00 | 8.1221E-01 | 1.5536E-04 | 1.9626E+00 | 3.2552E-06 | 1.00000 |
| 3.5 | 1.2990E+00 | 1.6198E+00 | 8.0746E-01 | 1.1289E-04 | 1.9477E+00 | 1.2615E-06 | 1.00000 |
| 4.0 | 8.8000E-01 | 1.6450E+00 | 8.0261E-01 | 8.6069E-05 | 1.9375E+00 | 6.6166E-07 | 1.00000 |
| 4.5 | 6.2372E-01 | 1.6669E+00 | 7.9767E-01 | 6.8023E-05 | 1.9306E+00 | 3.1377E-07 | 1.00000 |
| 5.0 | 4.5777E-01 | 1.6860E+00 | 7.9256E-01 | 5.5273E-05 | 1.9261E+00 | 1.8073E-07 | 1.00000 |
| 5.5 | 3.4570E-01 | 1.7029E+00 | 7.8811E-01 | 4.5930E-05 | 1.9237E+00 | 1.1285E-07 | 1.00000 |
| 6.0 | 2.6733E-01 | 1.7181E+00 | 7.8313E-01 | 3.8878E-05 | 1.9230E+00 | 7.1985E-08 | 1.00000 |

**Tab. A2** Values of the fit parameters, $\chi^2$/dof and $R^2$ for the fits of $a$ and $f$ as a function of $l$ with Eq. 4 of this work.

| Parameter | $p_0$ | $p_1$ | $p_2$ | $\chi^2$/dof | $R^2$ |
|---|---|---|---|---|---|
| $a$ | 5.1421E+01 | 2.9307E+00 | 3.2387E-01 | 1.3522E-05 | 1.00000 |
| $f$ | 1.2624E-03 | 1.9576E+00 | -4.7564E-01 | 1.0344E-12 | 1.00000 |

**Tab. A3** Values of the fit parameters, $\chi^2$/dof and $R^2$ for the fits of $b$, $c$ and $g$ as a function of $l$ with Eq. 5 of this work.

| Parameter | $q_0$ | $q_1$ | $q_2$ | $q_3$ | $\chi^2$/dof | $R^2$ |
|---|---|---|---|---|---|---|
| $b$ | 1.2273E+00 | 1.8829E-01 | -2.7291E-02 | 1.5923E-03 | 2.7766E-06 | 1.00000 |
| $c$ | 8.2992E-01 | -1.4232E-03 | -1.9297E-03 | 1.4504E-04 | 1.3002E-07 | 1.00000 |
| $g$ | 2.2239E+00 | -1.5055E-01 | 2.5665E-02 | -1.4905E-03 | 1.6580E-06 | 1.00000 |

**Tab A4** Values of the fit parameters, $\chi^2$/dof and $R^2$ for the fit $\delta(l)$ (Eq. 7) of the differences $S_{MC}(1,1,1,l)$ - $S(1,1,1,l)$ (Eq. 6) as a function of $l$.

| $A1$ | $t1$ | $A2$ | $t2$ | $y0$ | $\chi^2$/dof | $R^2$ |
|---|---|---|---|---|---|---|
| 6.4545e-04 | 7.3882e-01 | 8.2173e-02 | 2.8725e-01 | 8.1909e-07 | 9.2516e-10 | 1.00000 |